\documentclass[aps,pre,showpacs,twocolumn,groupedaddress]{revtex4}

\usepackage{graphicx,bm}
\begin{document}

\title{Isotropic-nematic transition in hard-rod fluids: relation between continuous and restricted-orientation models}
\author{Kostya Shundyak and Ren\'e van Roij}
\affiliation {Institute for Theoretical Physics, Utrecht University,\\
Leuvenlaan 4, 3584 CE Utrecht, The Netherlands}
\date{\today}

\begin{abstract}
We explore models of hard-rod fluids with a finite number of
allowed orientations, and construct their bulk phase diagrams
within Onsager's second virial theory. For a one-component fluid,
we show that the discretization of the orientations leads to the
existence of an artificial (almost) perfectly aligned nematic
phase, which coexists with the (physical) nematic phase if the
number of orientations is sufficiently large, or with the
isotropic phase if the number of orientations is small. Its
appearance correlates with the accuracy of sampling the nematic
orientation distribution within its typical opening angle. For a
binary mixture this artificial phase also exists, and a much
larger number of orientations is required to shift it to such high
densities that it does not interfere with the physical part of the
phase diagram.
\end{abstract}


\maketitle

\section{Introduction}
\label{introduct}Understanding the phase behavior of colloidal
suspensions of rodlike particles requires an accurate description
of their microscopic properties. Fluids of hard rods may be
considered as the simplest systems on which the models
incorporating particle orientational degrees of freedom can be
tested \cite{VL,F}. One of the exact theoretical results dates
back to Onsager \cite{O} who analyzed the transition from a
uniform isotropic phase to an orientationally ordered nematic
phase in a fluid of monodisperse hard needles. He realized that
the average pairwise (rod-rod) excluded volume is reduced in the
nematic phase compared to that in the isotropic phase, and argued
that the resulting gain of free volume (and hence translational
entropy) compensates the loss of orientation entropy (due to the
nematic ordering) at sufficiently high concentrations of rods
\cite{O}. Onsager derived a nonlinear integral equation for  the
orientation distribution function, a key quantity of the theory,
which is constant in the isotropic phase and peaked about the
director in the nematic phase. He circumvented the problem of
explicitly calculating the nematic orientation distribution
function (ODF) by adopting a variational Ansatz, which was
numerically checked to be rather accurate later \cite{VL}.

The generalization of the Onsager model to binary mixtures of rods
showed the possibility of strong fractionation
\cite{LCHDJCP84,OLJPC85} and even nematic-nematic demixing at
sufficiently high density, driven by a competition between
orientation entropy and ideal mixing entropy
\cite{BKP88,VLJPC93,RMJCP96}. The functional forms of ODF's in
these studies were either variational Gaussian
\cite{OLJPC85,VLJPC93}, truncated expansions in Legendre
polynomials \cite{LCHDJCP84}, or numerically determined on an
angular grid \cite{BKP88} or on a scaled angular grid
\cite{RMJCP96}. In all these cases the focus was on describing the
system with a continuum of orientations.

An alternative approach is to study models with a finite number
$N$ of allowed orientations while the positions of the centers of
mass of the rods remain continuous. The first such model was
proposed by Zwanzig \cite{Z}, with orientations of a rod to be
restricted to $N=3$ mutually perpendicular directions
$\bm{\hat{\omega}}_i, i=\{ 1,\ldots,N \}$. Despite its inability
to resolve the orientational structure of the one-particle
distribution function in any detail, it has been successfully
applied to explore wetting phenomena near a single hard wall and
in a slit \cite{RDEEL2000}, or phase diagrams of polydisperse
systems \cite{CCSSSJCP2000,MRCJCP2003}. The main advantage of such
discrete models in comparison with the continuous ones is their
computational simplicity. The combination of spatial inhomogeneity
and/or polydispersity with a continuum of orientations is rather
demanding numerically, and the computational efforts can be
reduced significantly by discretizing the orientations
\cite{CCSSSJCP2000,CN92,SRJPCM2001}. The hope has, of course,
always been that with an increase of the number of allowed
orientations one would smoothly approach the continuum limit. Here
we show that this is {\it not} the case.

The possibility of a continuous interpolation between results for
the discrete models on the one hand and Onsager-like solutions on
the other has first been questioned by Straley \cite{SJCP72} in
studies of models with dodecahedral ($N=6$) and icosahedral
($N=10$) symmetries. He concluded that they do not trend towards
the continuum solution due to the single allowed orientation
within the typical opening angle ($\approx \pi/9$) of the nematic
distribution at coexistence. Unfortunately, one cannot proceed the
sequence of models with $N=3,6,10$ any further, since a larger
fully symmetric set of orientations on the unit sphere does not
exist. In order to be able to study the effect of discretizing the
allowed orientations we give up part of the symmetry of the set,
and this allows us to connect continuous and discrete models. We
apply our method not only to a one-component system of rods, but
also to binary mixtures, which may be considered as the simplest
polydisperse systems.

This paper is organized as follows. In Sec. \ref{mono} we derive
the grand potential functional for a model with discrete number of
allowed rod orientations from the Onsager functional. We calculate
bulk equations of state for specific orientational sets, and
determine the number of orientations required to resemble the
continuous Onsager solution. In Sec. \ref{binary} we apply the
method to construct bulk phase diagrams of binary mixtures of thin
and thick rods. We demonstrate that their structure can be
significantly modified due to orientational discretization. A
summary and some discussion of our results will be presented in
Sec. \ref{discussion}.

\section{Monodisperse rods}
\label{mono} Consider a fluid of hard rods of length $L$ and
diameter $D$ ($L\gg D$) in a macroscopic volume $V$ at temperature
$T$ and chemical potential $\mu$.  The ``Onsager'' grand potential
functional $\Omega[\rho]$ of the one-particle distribution
function $\rho(\bm{\hat{\omega}})$ can be written, within second
virial approximation,  as \cite{O}
\begin{eqnarray}
\label{Omegacont} \beta\Omega[\rho(\bm{\hat{\omega}})]&=& \int d
\bm{\hat{\omega}} \rho(\bm{\hat{\omega}}) \Big(\ln [\rho
(\bm{\hat{\omega}}) \nu]-1-\beta\mu \Big)\nonumber\\
&&+ \frac{1}{2}\int d\bm{\hat{\omega}} d\bm{\hat{\omega}}'
E(\bm{\hat{\omega}},\bm{\hat{\omega}}')
\rho(\bm{\hat{\omega}})\rho(\bm{\hat{\omega}}'),
\end{eqnarray}
where $\beta=(kT)^{-1}$ is the inverse temperature, $\nu$ is the
rod's thermal volume, and
$E(\bm{\hat{\omega}},\bm{\hat{\omega}}')$ is the excluded volume
of rods with orientations $\bm{\hat{\omega}}$ and
$\bm{\hat{\omega}}'$. The function $\rho(\bm{\hat{\omega}})$ is
normalized as $n=\int d\bm{\hat{\omega}} \rho(\bm{\hat{\omega}})$,
with $n$ the bulk number density (which depends on $\beta\mu$).
The minimum condition $\delta\Omega[\rho(\bm{\hat{\omega}})]/
\delta\rho(\bm{\hat{\omega}})=0$ on the functional leads to the
nonlinear integral equation
\begin{eqnarray}
\label{nonlinsetcont} \ln[\rho(\bm{\hat{\omega}})\nu] +\int
d\bm{\hat{\omega}}' E(\bm{\hat{\omega}},\bm{\hat{\omega}}')
\rho(\bm{\hat{\omega}}') =\beta\mu,
\end{eqnarray}
to be solved for the equilibrium distribution
$\rho(\bm{\hat{\omega}})$.

Models with a discrete number $N$ of allowed rod orientations can
be systematically derived from the continuous model
(\ref{Omegacont}) by dividing the unit sphere into solid sectors
$\Delta\bm{\hat{\omega}}_i, (i=1,\ldots,N)$ around vectors
$\bm{\hat{\omega}}_i$, and fixing the rod density
$\rho(\bm{\hat{\omega}})=\rho(\bm{\hat{\omega}}_i)$ within each
sector as well as the excluded volume
$E(\bm{\hat{\omega}},\bm{\hat{\omega}}')=
E(\bm{\hat{\omega}}_i,\bm{\hat{\omega}}_j)$ for every pair of
sectors. The grand potential functional $\Omega[\rho_i]$ of such
"orientationally discretized" fluid with the density
$\rho_i=\rho(\bm{\hat{\omega}}_i) \Delta\bm{\hat{\omega}}_i$ and
the excluded volumes $E_{ij}=
E(\bm{\hat{\omega}}_i,\bm{\hat{\omega}}_j)$ is
\begin{eqnarray}
\label{Omegadiscr} \beta\Omega[\rho_i]&=&\sum_{i=1}^N \rho_i
\Big(\ln [\rho_i \nu]-1-\beta\mu \Big)+\frac{1}{2}\sum_{i,j=1}^N
E_{ij} \rho_i \rho_j \nonumber\\
&&-\sum_{i=1}^N \rho_i \ln \Delta\bm{\hat{\omega}}_i
\end{eqnarray}
with normalization $n=\sum_{i=1}^N \rho_i$. The last term in Eq.
(\ref{Omegadiscr}) represents the contribution due to the
discretization procedure into the grand potential $\Omega$, i.e.
the intrinsic difference between continuous and discrete models.
For a homogeneous distribution of vectors $\bm{\hat{\omega}}_i$ on
the unit sphere and $\Delta \bm{\hat{\omega}}_i=\Delta
\bm{\hat{\omega}}$ (i.e. for the models with $N=3,6$ and $10$
\cite{SJCP72}), it trivially shifts the chemical potential
$\beta\mu_d=\beta\mu +\ln\Delta \bm{\hat{\omega}}$, which does not
have any consequence for the solutions $\rho_i$ at a fixed $n$,
and for the thermodynamics of the isotropic-nematic transition.
However, when $\Delta\bm{\hat{\omega}}_i$ is not the same for all
$i$, it acts as an external orientational field that tends to
favor the larger sectors over the smaller ones. This becomes
explicit if we consider the Euler-Lagrange equations that
corresponds to the discrete functional (or equivalently the analog
of Eqs. (\ref{nonlinsetcont}))
\begin{eqnarray}
\label{nonlinsetdiscr} \ln[\rho_i \nu] +\sum_{j=1}^N E_{ij} \rho_j
=\beta\mu+\ln\Delta \bm{\hat{\omega}}_i,
\end{eqnarray}
now to be solved for $\rho_i$. Note that the equation of state
$p=p(n,T)$ does not pick up an additional term from
discretization,
\begin{eqnarray}
\label{pdiscr} \beta p=n+\frac{1}{2}\sum_{j=1}^N E_{ij} \rho_i
\rho_j,
\end{eqnarray}
but the distributions $\rho_i$ to be inserted into it {\em do}
depend on the discretization.

Further discussion requires a specification of the set of allowed
orientations $\bm{\hat{\omega}}_i, i=1,\ldots,N$ and the
associated solid sectors $\Delta\bm{\hat{\omega}}_i$.
Unfortunately, it is impossible to completely cover a surface of
the unit sphere by equal regular spherical $M$-polygons, where $M$
indicates the number of polygon's sides (only $5$ Platonic solids
exist). But symmetries of the function $\rho(\bm{\hat{\omega}})$
can be explicitly included into the set of vectors
$\bm{\hat{\omega}}_i$ in order to simplify the problem. For the
present study we fix the $\bm{\hat{z}}$ axis of the coordinate
system to be parallel to the nematic director $\bm{\hat{n}}$ and
assume uniaxial symmetry of the function
$\rho(\bm{\hat{\omega}})=\rho(\theta)$, with
$\theta=\arccos(\bm{\hat{\omega}} \cdot\bm{\hat{n}})$ the angle
between $\bm{\hat{\omega}}$ and $\bm{\hat{n}}$. The azimuthal
angle is denoted by $\phi$, and hence we characterize a vector
$\bm{\hat{\omega}}=(\sin\theta\cos\phi,\sin\theta\sin\phi,\cos\theta)$
by the angles $\theta$ and $\phi$. The ``up-down" symmetry of the
nematic phase reduces the orientational space to half the
upper-hemisphere, i.e. $\theta\in[0,\pi/2]$ and $\phi\in[0,\pi]$.
As we do not expect any azimuthal symmetry breaking we restrict
attention to $N_{\phi}$ uniformly distributed values for $\phi$
for every allowed $\theta$. We have considered a uniform
distribution of $N_{\theta}$ polar angles $\theta\in[0,\pi/2]$,
i.e.
\begin{eqnarray}
\label{orientvectors1} (\theta_k,\phi_l)=\Big(\frac{\pi
(k-1)}{2(N_{\theta}-1)},\frac{\pi
(l-1)}{N_{\phi}-1}\Big),\nonumber\\
k=1,\ldots,N_{\theta}, \;\;\;\; l=1,\ldots,N_{\phi},
\end{eqnarray}
as well as a uniform distribution of $N_{\theta}$ values of
$\cos\theta \in[0,1]$, i.e.
\begin{eqnarray}
\label{orientvectors2}
(\theta_k,\phi_l)=\Big(\arccos\Big[1-\frac{k-1}{N_{\theta}-1}\Big],
\frac{\pi l-1}{N_{\phi}-1}\Big),\nonumber\\
k=1,\ldots,N_{\theta}, \;\;\;\; l=1,\ldots,N_{\phi}
\end{eqnarray}
with the conventional definition of
$\Delta\bm{\hat{\omega}}_i=\int_{\Omega_i} \sin\theta d\theta
d\phi$.

Figure \ref{eqstate} shows the dimensionless pressure $p^*=\beta p
L^2D$ as a function of the dimensionless bulk density $n^*=nL^2D$
for the grid (\ref{orientvectors1}) with different $N_{\theta}$
and $N_{\phi}=5$. The plateaux (of the solid lines) correspond to
the isotropic-nematic coexistence, obtained by equating pressure
and chemical potential in the two phases. For $N_{\theta}\leq 9$
the transition occurs between the isotropic phase ($I$) and an
(almost) perfectly aligned nematic phase ($A$) with the pressure
being close to the the ideal gas pressure. Note that such grids
correspond to a single $\theta$ within the "Onsager" opening
angle, i.e. $0\leq \theta_1 \leq\pi /18<\theta_2$. As soon as
$0\leq \theta_1 < \theta_2\leq\pi /18$, or $N_{\theta}>9$, the
distribution function $\rho(\theta)$ at isotropic-nematic
coexistence starts to converge to the continuous solution. These
results are in full agreement with the previous explanation of
Straley \cite{SJCP72}. Equations of state for the model
(\ref{orientvectors2}) are very similar to Fig. \ref{eqstate} but
start to resemble Onsager-like distribution function for
$N_{\theta}>80$ due to the poor sampling of $\bm{\hat{\omega}}_i$
near the nematic director. Equations of state for the Zwanzig
($N=3$), dodecahedral ($N=6$) and icosahedral ($N=10$) models were
calculated using the original formulations \cite{SJCP72}, and are
included for comparison. Our results for $N_{\theta} \leq 9$ seem
to converge well to these existing results.

For $N_{\theta}>20$ the pressure of the high-density nematic phase
clearly demonstrates a linear dependence on the bulk density, i.e.
$\beta p(n)\sim n$. With increasing $N_{\theta}$ it gradually
approaches a limiting scaling behavior $\beta p(n)=3n$,
established for the continuous Onsager solution by means of a
scaling argument \cite{RMEL96}.

The discretization of the rod's allowed orientations shows the
existence of an "artificial transition" from a less-ordered
nematic phase ($N$) to a near-perfectly aligned phase ($A$) for
$N_{\theta}>9$, as indicated by the dashed horizontal lines in
Fig.\ref{eqstate}. It occurs due to the same competition between
excluded volume and orientational entropy as in the IN transition,
and puts an additional constraint on the description of the
nematic bulk state by restricted-orientation models. Below we
argue that it has important consequences for discrete models of
(polydisperse) hard-rod mixtures, where separation into nematic
phases with different composition occurs at sufficiently high
densities.

\section{Binary mixtures}
\label{binary} Consider a binary mixture of thin ($D_1$) and thick
($D_2$) hard rods of equal length $L$ and the diameter ratio
$d=D_2/D_1=4$ in a macroscopic volume $V$ at temperature $T$ and
chemical potentials $\mu_1$ and $\mu_2$, respectively. The
``Onsager'' grand potential functional for this system can be
written as \cite{VL}
\begin{eqnarray}
\label{Omegacontbin} \beta\Omega[\{
\rho_{\sigma}(\bm{\hat{\omega}})\}]= \sum_{\sigma=1}^2 \int d
\bm{\hat{\omega}} \rho_{\sigma}(\bm{\hat{\omega}}) \Big(\ln
[\rho_{\sigma} (\bm{\hat{\omega}}) \nu]-1-\beta\mu_{\sigma}
\Big)\nonumber\\
+ \frac{1}{2}\sum_{\sigma,\sigma'=1}^2 \int d\bm{\hat{\omega}}
d\bm{\hat{\omega}}'
E_{\sigma,\sigma'}(\bm{\hat{\omega}};\bm{\hat{\omega}}')
\rho_{\sigma}(\bm{\hat{\omega}})\rho_{\sigma'}(\bm{\hat{\omega}}'),
\end{eqnarray}
with normalization $n_{\sigma}=\int d \bm{\hat{\omega}}
\rho_{\sigma}(\bm{\hat{\omega}})$. It is known from previous work
\cite{RMDP98,SRPRL02,SRPRE03} that the bulk phase diagram of this
system exhibits (i) strong fractionation at isotropic-nematic
($I-N_2$) coexistence, (ii) nematic-nematic ($N_1-N_2$)
coexistence ending in a consolute $N_1N_2$ point at sufficiently
high pressure, and (iii) an $IN_1N_2$ triple point. The discrete
version of this model follows directly from (\ref{Omegadiscr}) as
\begin{eqnarray}
\label{Omegadiscrbin} \beta\Omega[\{\rho_{\sigma
i}\}]=\sum_{\sigma=1}^2 \sum_{i=1}^N \rho_{\sigma i} \Big(\ln
[\rho_{\sigma i} \nu_{\sigma}]-1-\beta\mu_{\sigma} \Big)
\nonumber\\ +\frac{1}{2}\sum_{\sigma \sigma'=1}^2\sum_{i,j=1}^N
E_{\sigma i; \sigma' j} \rho_{\sigma i} \rho_{\sigma' j}
-\sum_{\sigma=1}^2 \sum_{i=1}^N \rho_{\sigma i} \ln
\Delta\bm{\hat{\omega}}_i
\end{eqnarray}
with the densities $\rho_{\sigma
i}=\rho_{\sigma}(\bm{\hat{\omega}}_i) \Delta\bm{\hat{\omega}}_i$,
the excluded volumes $E_{\sigma i;\sigma' j}= E_{\sigma \sigma'}
(\bm{\hat{\omega}}_i, \bm{\hat{\omega}}_j)$ and the number
densities normalization $n_{\sigma}=\sum_{i=1}^N \rho_{\sigma i}$.
Figure \ref{eqstatebin} shows the phase diagrams for discrete
systems in the $p^*-x$ representation with the dimensionless
pressure $p^*=\beta p L^2D_1$ and the mole fraction of thick rods
$x=n_2/(n_1+n_2)$, for several orientational grids
(\ref{orientvectors1}) with $N_{\theta}=11$ (a), $20$ (b), $30$
(c), $50$ (d) and $N_{\phi}=10$. Note that all four
discretizations are such that they reproduce the physical
``Onsager''-like $I-N$ transition at $x=0$ and $x=1$, at pressures
$p^* \approx 17.7$ and $4.4$, respectively. However, the existence
of the artificial aligned nematic phase $A$ gives rise to spurious
$I-A$, $N_1-A$ and $N_2-A$ phase equilibria, where $I$, $N_1$ and
$N_2$ are the physical isotropic and nematic phases. (For the
coarsest discretization with $N_{\theta}=11$
(Fig.\ref{eqstatebin}(a)) the $N_1$ phase is stable in a very
narrow region beyond the resolution of the picture.) Upon refining
the discretization from $N_{\theta}=11$ the $IN_2A$ triple point
($\nabla$) shifts to higher pressures, and combines with the
$IN_1A$ triple point ($\triangle$) at $N_{\theta}=30$
(Fig\ref{eqstatebin}(c)) to form the physical $IN_1N_2$ triple
point (now denoted by $\nabla$) and an artificial $N_1N_2A$ triple
point ($\triangle$) at slightly higher pressure. Further grid
refinements to $N_{\theta}=50$ yields the physical phase diagram
with an $IN_1N_2$ triple point and $N_1N_2$ consolute point as in
Ref.\cite{RMDP98}, but with a spurious $N-A$ coexistence at
pressures beyond the ``physical'' part of the phase diagram. At
these pressures one does not distinguish $N_1$ and $N_2$ nematic
phases, but the single nematic phase denoted by $N$ here.

\section{Summary}
\label{discussion} We have explored the connections between
continuous and restricted orientation models of monodisperse and
binary hard-rod fluids (in the Onsager ``needle'' limit $L\gg D$).
Our main finding is that a discretization of the orientations
leads to existence of a nonphysical almost perfectly aligned
nematic phase ($A$) at high densities. If the discretization is
coarse, i.e., the number of allowed orientations is small, then
the $A$ phase can coexist with the isotropic phase ($I$), and at
sufficiently fine discretization with the nematic phase $N$. We
also found that the continuum limit requires a finer orientation
grid for a mixture than the one-component fluid.

Clearly, our results are strongly influenced by the adopted limit
$L \gg D$, but we would expect similar effects, although weaker,
for finite $L/D$ ratio.

These findings could be relevant for the study of inhomogeneous
and/or polydisperse fluids of rods, which are computationally more
demanding and hence impose the use of rather coarse grid of
orientations in order to be tractable and practical. The present
study shows that care must be taken with such rather coarse grids,
since they can give rise to an artificial, discretization-induced
aligned nematic phase.

\acknowledgements We would like to thank Prof. J.P. Straley for
useful correspondence. This work is part of the research program
of the ``Stichting voor Fundamenteel Onderzoek der Materie
(FOM)'', which is financially supported by the ``Nederlandse
organisatie voor Wetenschappelijk Onderzoek (NWO)''.

\newpage
FIGURE CAPTIONS
\begin{enumerate}
\item Equation of state for models with different number of
allowed polar angles $N_{\theta}$. Positions of the phase
transitions are indicated by the horizontal lines. The continuous
Onsager solution (Ons) can be reproduced in the present density
interval with $N_{\theta}\geq 50$. Dotted lines correspond to
equations of state for the Zwanzig (Z), dodecahedral (6S) and
icosahedral (10S) models \cite{SJCP72}. The dashed horizontal
lines correspond to a spurious nematic-nematic ($N-A$) transition
due to poor discretization of the allowed orientations.

\item Bulk phase diagrams of a binary thin-thick mixture of hard
rods (diameter ratio $D_2/D_1=4.0$, equal length $L\gg D_2$), in
the pressure-composition representation, with $p^*$ the
dimensionless pressure and $x$ the mole fraction of the thicker
rods. We distinguish the low-pressure isotropic phase ($I$),
high-pressure nematic phases ($N_1$ and $N_2$), aligned phase $A$,
upper ($\triangle$) and lower ($\nabla$) triple phase coexistence
and an $N_1-N_2$ critical point ($\ast$). The grey regions,
enclosed by the binodals, denote the two-phase regimes, and the
tie lines that connect coexisting phases, are horizontal.

\end{enumerate}

\begin{widetext}

\newpage
\begin{figure}[h]\centering
\includegraphics[width=8.5cm]{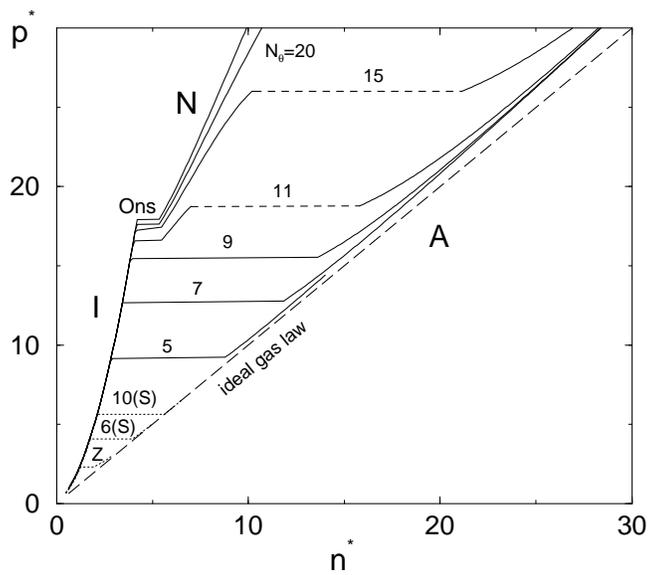}
\caption{\label{eqstate} K. Shundyak and R. van Roij}
\end{figure}

\begin{figure}[h]
\includegraphics[scale=0.4]{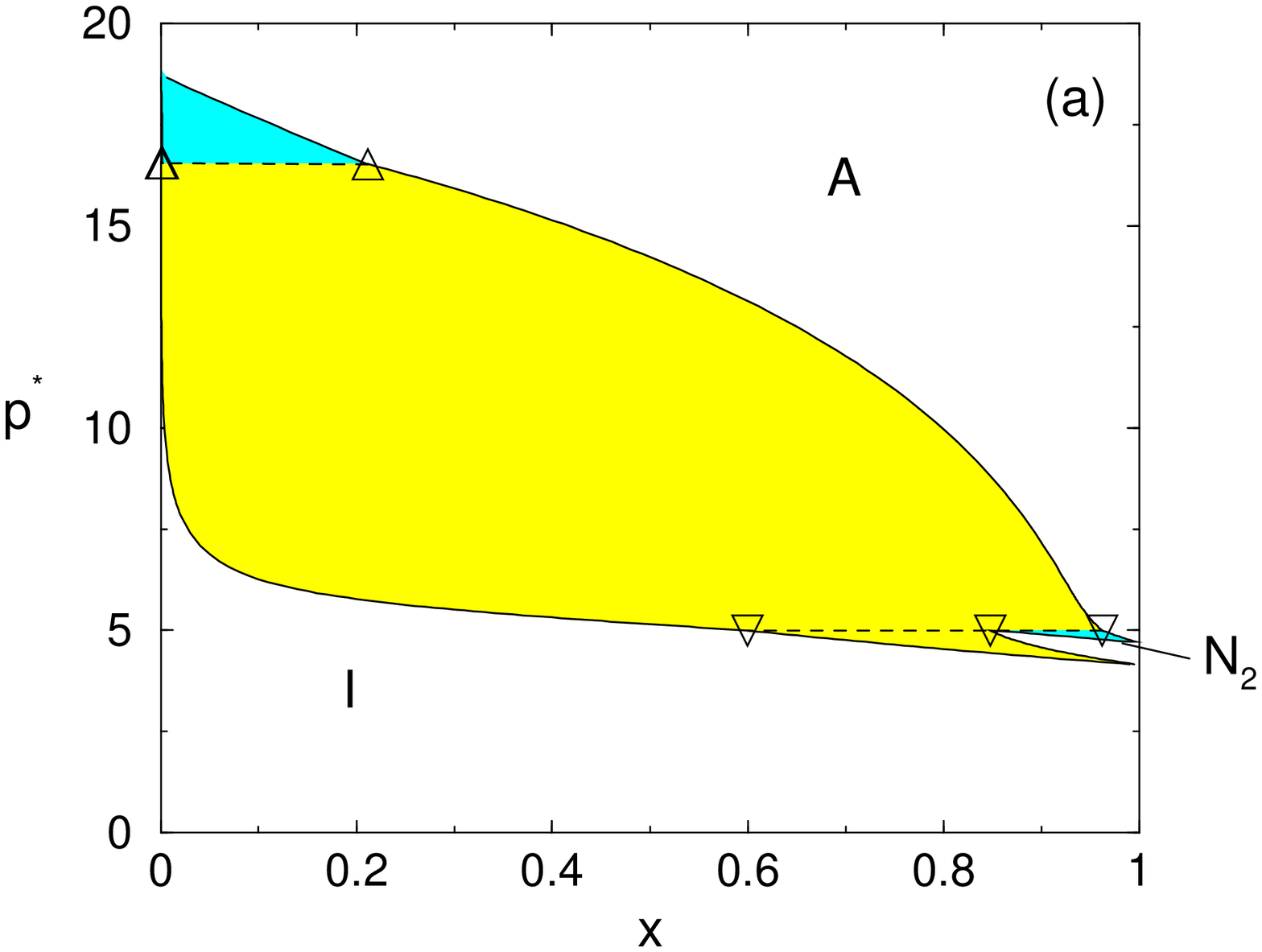}
\includegraphics[scale=0.4]{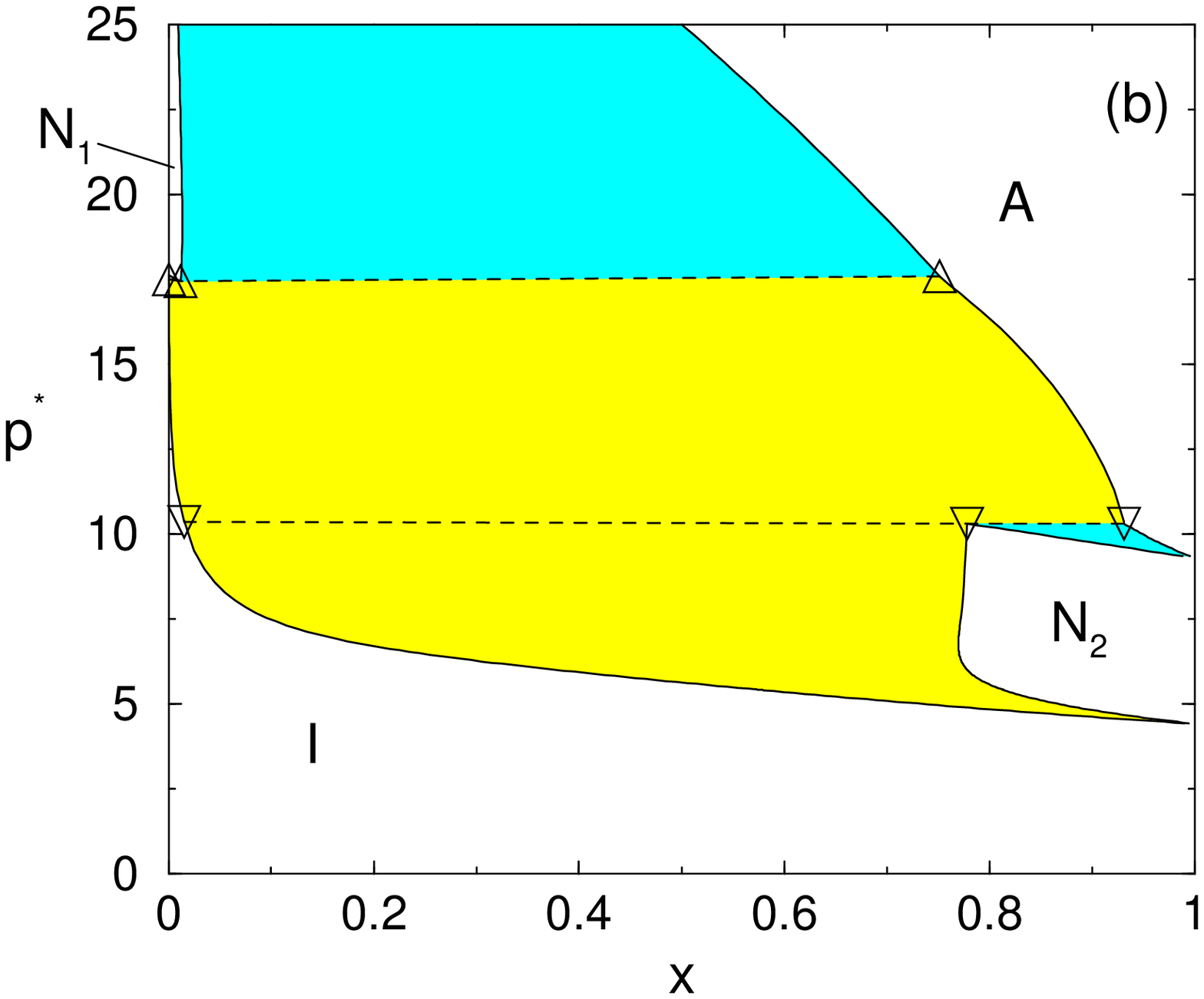}
\includegraphics[scale=0.4]{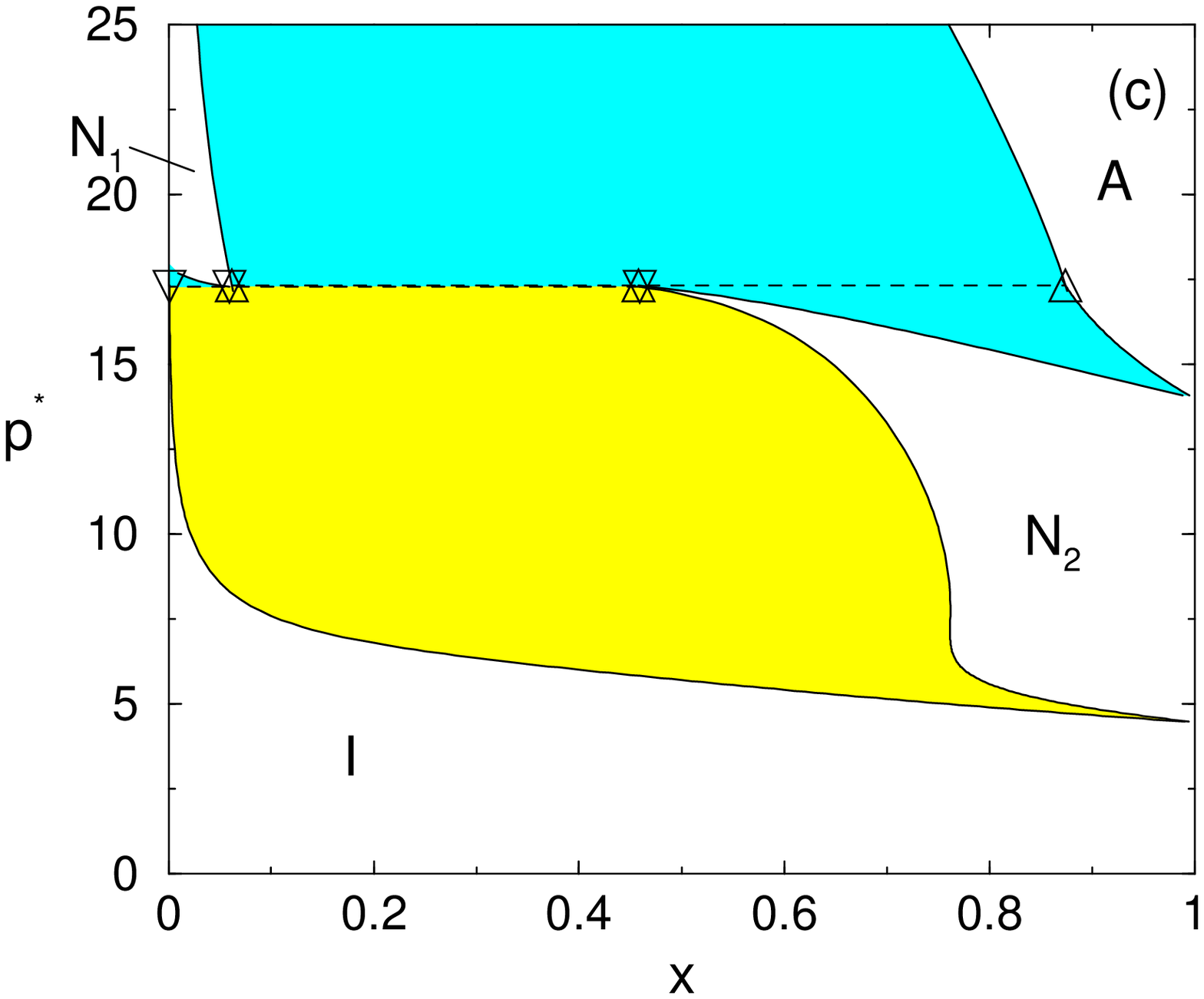}
\includegraphics[scale=0.4]{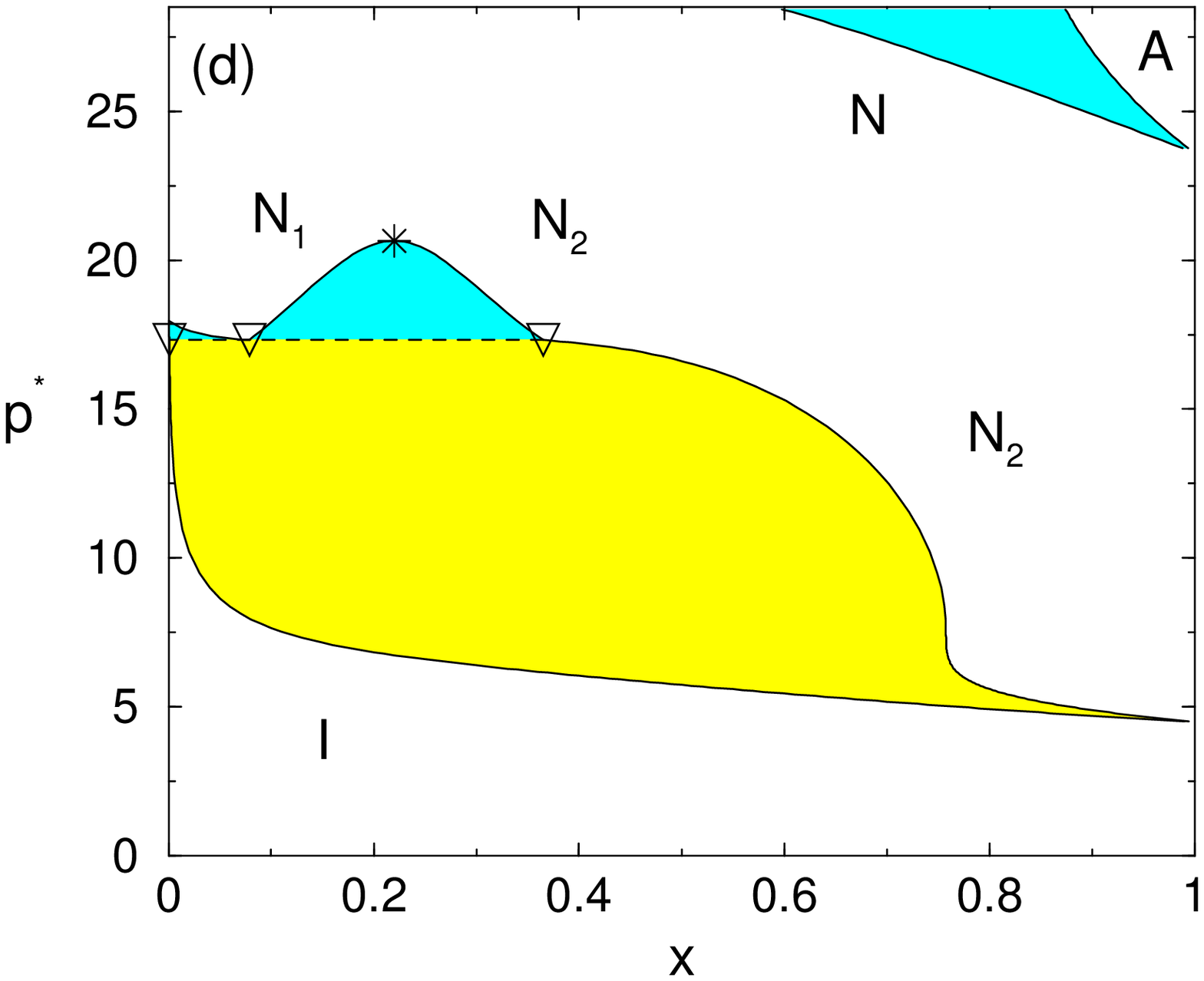}
\caption{\label{eqstatebin} K. Shundyak and R. van Roij}
\end{figure}

\end{widetext}

\end{document}